\documentclass[aps.pre,showpacs]{revtex4}
\usepackage{graphicx}
\usepackage{color}\def\be{\begin{equation}}

\def\ee{\end{equation}}
\def\bfi{\begin{figure}}
\def\efi{\end{figure}}
\def\bea{\begin{eqnarray}}
\def\eea{\end{eqnarray}}

\begin{document}

\title{Development and regression of a large fluctuation}

\author{Federico Corberi}

\affiliation{Dipartimento di Fisica ``E.~R. Caianiello'', and INFN,
Gruppo Collegato di Salerno, and CNISM, Unit\`a di Salerno,Universit\`a  di Salerno,
via Giovanni Paolo II 132, 84084 Fisciano (SA), Italy.}

\begin{abstract}
{We study the evolution leading to (or regressing from) a large fluctuation
in a Statistical Mechanical system. We introduce and study analytically a simple model of many
identically and independently distributed microscopic variables $n_m$
($m=1,M$)
evolving by means of a master equation.
We show that the process
producing a non-typical fluctuation with a value
of $N=\sum _{m=1}^Mn_m$ well above the average
$\langle N\rangle$ is slow. Such process is characterized by
the power-law growth of the largest possible observable value of $N$
at a given time $t$. We find similar features also for
the reverse process of the regression from a rare state with
$N\gg \langle N\rangle$ to a typical one with $N \simeq \langle N\rangle$.}
\end{abstract}

\maketitle

\section{Introduction}

The occurrence of fluctuations is at the heart of most physical phenomena \cite{LL}.
Typically, in an extended system made of a large number $M$ of microscopic constituents,
like those usually considered in equilibrium thermodynamics, a collective variable  $N$
(like the particle number or the energy) evolves as to stay most of the time
close to its average value $\langle N\rangle $. Large deviations
are rare and become progressively less frequent as
$N$ moves away from the average.
For this reason they are neglected in many practical applications.
However in some cases they can have important consequences.
This happens, for instance,
when their occurrence leads the system to an absorbing state, namely a configuration
that cannot be escaped \cite{Hinrichsen2000}. Examples include the extinction of a species, the failure of a device,
or the bankruptcy of a company. The latter indeed was the first problem for which the rigorous results of large deviation theory were applied \cite{Cramer}.
In addition, large deviations play a
prominent role in many non-equilibrium phenomena, e.g. in the decay of
metastable states \cite{langer}.

Configurations corresponding to a large fluctuation
are usually very different from those typically observed when
$N \simeq \langle N \rangle$ (to ease the notation we use the same symbol for 
the stochastic variable and its possible outcomes), because the system explores
a seldom visited region of phase-space that may have peculiar properties.
Hence, the question arises of how the representative point moves
to reach such low-probability sectors, or, in other words, what are the properties of
the dynamical process producing a large deviation.
This issue is not only an important and largely unexplored topic in large deviation theory,
but might also represent a first step towards the detection and control
of fluctuations, with important applications concerning the
predictability of catastrophic events. Likewise, the reverse process, whereby the typical behavior is
recovered after a rare event, has also theoretical and practical interest.

In this paper we study such problems in a simple but sufficiently general model where
$N=\sum _{m=1}^M n_m$ is the sum of a large number $M$ of
independent variables $n_m=0,1,\dots$ identically distributed with probability $p(n_m)$.
The creation of a fluctuation
is studied by evolving the initial distribution
$p(n,0)=p(n_m=n,t=0)$ of the microvariables $n_m$ in a typical state of the system
with $N=\langle N \rangle$, until a deviation
with $N \neq \langle N \rangle$
is observed. Solving the master equation yields the evolution of the probability
$P(N,M,t)$ that the collective variable takes a given
value $N$. This quantity provides a detailed description
of the whole fluctuation spectrum of $N$,
characterizes the event whereby
the fluctuation is built, and identifies its relevant properties.
In the same way, one can study the disruption of a large deviation by studying
the evolution from an initial condition with
$N \neq \langle N \rangle$.

In the present Article we choose the master equation governing the dynamics of $p(n,t)$
such as to have the stationary solution
\be
p_{st}(n)\propto (n+1)^{-k}.
\label{statpsmall}
\ee
Systems with fat tail distributions analogous to the one considered here are found in
natural sciences, social sciences, and economics. Among many examples we can mention
the magnitude of earthquakes \cite{equakes}, the spreading of forest fires \cite{ffires},
rain events \cite{revents}, size of cities \cite{csize},
wealth distribution \cite{eqwealth}, price returns of stock's indices \cite{preturn},
degree distribution of networks \cite{barab}.

Our choice of $p_{st}$ is not only motivated by its ubiquitous character, but also stems
from general considerations regarding the actual probability to observe
large fluctuations. As we discuss below, such probability is particularly large
for the model under consideration.

We have already mentioned that in extended systems
the large deviations of a collective variable $N$
are generally strongly suppressed. Indeed the probability
$P(N,M)$ usually obeys \cite{touch} the following {\em large deviation principle} (LDP):
\be
\lim _{M\to \infty}\ln P(N,M)=-MR(\rho) \; ,
\label{ldp}
\ee
where $\rho =\frac{N}{M}$ and $R$ is the {\it rate function}. For simplicity, in the above equation and in the following we omit the
dependence on $t$.
Eq. (\ref{ldp}) implies that $P(N,M)$ is always exponentially small (in $M$) except for the values
of the {\em density} $\rho $ for which $R(\rho)=0$. Notice that at least one
such configuration is bound to exist in order to preserve
normalization of probability as $M \to \infty$.
The simplest case is when there is a single value of
$\rho $ yielding $R(\rho)=0$. This value trivially coincides with the average
$\langle \rho \rangle =\frac{\langle N\rangle}{M}$ for large $M$.
In this scenario, the outcome of a measurement
is almost always close to $\langle \rho \rangle $, whereas
sizable fluctuations are extremely rare and can only be
observed in systems of mesoscopic scale, namely with $M$ not too large.
The same mathematical structure applies
to all those cases where $N$ is formed by the addition
of many microscopic contributions. This is nicely illustrated by the much studied \cite{current} problem of
the fluctuations of a {\it charge} $N=\int _0 ^\tau j(t) dt$, $j$ being a
current, flowing in a certain time interval $\tau$ through a system.

Despite all of the above, there are important cases where large fluctuations are not exponentially suppressed as
in Eq. (\ref{ldp}), because the LDP breaks down for some range of values of $N$.
In this situation, $R(\rho)$ can vanish
not only for one (or some isolated) specific value(s) of $\rho$, but in a whole
interval $\rho \in \backslash \hspace{-.2cm}{\cal I}$:
\be
R(\rho )\equiv 0, \hspace{1cm} \mbox{for} \hspace{.5cm}\rho \in \backslash \hspace{-.2cm}{\cal I},
\label{condnoldp}
\ee
In this case Eq. (\ref{ldp}) must more properly be re-written as
\be
\ln P(N,M)\simeq -M\left [R(\rho ) + {\cal R}(M,\rho)\right ], \hspace{1cm}\mbox{for large }M.
\label{ldpbreak}
\ee
When the LDP holds, $R$ is finite and the second term on the r.h.s. of the equation above
is a correction to the leading behavior for finite $M$: $\lim _{M\to \infty} \frac{{\cal R}(M,\rho)}{R(\rho)}=0$.
Then, in the large-$M$ limit,
Eq. (\ref{ldp}) is meaningful, since fluctuations are fully described by $R$.
However, if Eq. (\ref{condnoldp}) holds, then ${\cal R}$ becomes the relevant term and Eq. (\ref{ldp})
is useless.
This implies that fluctuations are reduced more softly in $\backslash \hspace{-.2cm}{\cal I}$ than
where Eq. (\ref{ldp}) holds and, depending on the system, significant deviations
may have a good chance to form and be detected.

Situations where Eq. (\ref{condnoldp}) holds
are observed in a variety of systems.
A first notable example is represented by magnetic materials where, below the critical temperature, the probability
distribution of the (fluctuating) spontaneous magnetization $N$ exhibits \cite{magnetic_no_ldp}
a structure like the one in Eq. (\ref{ldpbreak}),
where $M$ is the number of spins. In this
case $R$ vanishes in the whole region $\backslash \hspace{-.2cm}{\cal I}$ such that $\rho \in [-\rho _0,\rho _0]$,
where $\rho _0$ is the absolute value of the
(average) spontaneous magnetization (per spin).
Another example is the case of a Brownian walker on a line, with hopping rates
retaining memory of the previous history \cite{browninan_memory}. The probability of moving a certain
distance $N$ after $M$ steps takes a form like the one of Eq. (\ref{ldpbreak}). For instance, for a particular choice of the
memory term, $R$ vanishes in the entire region $\rho >0$, i.e. when the particle,
starting from the origin, moves to the right. Other examples include fluctuations of driven Maxwell-Lorentz particles \cite{grad},
quantum quenches \cite{gamba}, disordered systems \cite{disordered}, and many others \cite{other_ldp_break}.

As already anticipated, the model that we will study in this Article is conceived in such a way that the
stationary probability $P_{st}(N,M)$ obeys the crucial LDP-breaking property Eq. (\ref{condnoldp}).
Indeed, it is well known \cite{corberijpa} that assuming the distribution Eq. (\ref{statpsmall})
with $k>2$, it follows that $-\lim _{M\to \infty}M^{-1}\ln P_{st}(N,M)= R_{st}(\rho )=0$ in the entire range
$\backslash \hspace{-.2cm}{\cal I}\equiv \{\rho\, \vert \,\rho>\langle \rho \rangle \}$.
Correspondingly, the LDP Eq. (\ref{ldp}) does not apply in that region (notice that the same phenomenon
does not occur for $k\le 2$, see the discussion in Sec. \ref{stat}).
As a consequence, the condition for the
observability (in the sense discussed above, after Eq. (\ref{condnoldp}))
of large deviations is met in the interval
$\rho \in \backslash \hspace{-.2cm}{\cal I}$.
The detailed theoretical study of the formation and regression of a large deviation that we will carry out in the present Article
for an analytically tractable model could therefore pave the way to the experimental investigation of such processes.
Moreover, the model studied here could represent a simple paradigm for a
class of systems, like those  mentioned above, where the condition (\ref{condnoldp}) is satisfied.
For example, in the previously discussed magnetic context,
the problem at hand would correspond to study the spontaneous process
whereby a typical configuration with $\rho =\rho _0$
evolves to a probabilistically unfavored one with $\rho < \rho _0$
due to thermal fluctuations, and the regression
to the initial state.

Probabilistic setups, similar in spirit to the one introduced and studied in the present work,
have been already considered in models
of simplicial quantum gravity \cite{bialas1,bialas2},
non-equilibrium driven systems \cite{zrp},
L\'evy walks \cite{levy}, and other systems \cite{others} (see Sec. \ref{themodel} for
more details). However, to the best of our knowledge, the process whereby fluctuations form and regress
has never been previously investigated.

Due to the analytical tractability of the model, we can derive several significant
results from its solution.
Firstly, the nature of the process associated with the production and
regression of fluctuations is radically different if
it occurs in the region
${\cal I}\equiv \{\rho\, \vert \,\rho\le \langle \rho \rangle \}$
where the LDP holds, or in $\backslash \hspace{-.2cm}{\cal I}$, where it is violated.
In the former case the evolution is fast and relatively simple, with an exponential
convergence towards the stationary form. In the latter,
it displays a slow non-trivial evolution.
This happens because the mechanism whereby $P(N,M,t)$
approaches $P_{st}(N,M)$ is only effective up to a typical finite fluctuation scale
$N\sim \nu (t)$, while larger values of $N$ are left untouched.
The characteristic value $\nu (t)$ increases slowly
in an algebraic way.
This leads to an everlasting 
aging phenomenon, which closely resembles the dynamics of systems
crossing a phase transition \cite{phtrans}. The probability $P(N,M,t)$
attains the stationary value for increasingly large values
of $N$. However, at each time there will always exist a sufficiently large
value of $N$ beyond which stationarity is not reached.

Secondly, considering the growth of spontaneous fluctuations,
the mechanism whereby LDP breaks down, starting from an initial configuration that
satisfies it, shows a nontrivial
interplay between $M$ and $t$. Specifically, while violations of LDP are enhanced
at large times, as expected, they are reduced by increasing $M$.
This shows that, when studying the large time behavior of a fluctuating system,
attention should be paid to the order of the limits $\lim _{M\to \infty}$ and $\lim _{t\to \infty}$,
which again reminds the physics of phase transitions.

This paper is organized as follows: In Sec. \ref{themodel} we introduce the
statistical model and set the notations.
In Secs. \ref{stat} we discuss the properties of the model in the
large-time limit when stationarity is reached. The breakdown of LDP is discussed and
related to a {\it condensation} phenomenon.
In Sec. \ref{dynam} the formation (Sec. \ref{p_crea}) and suppression (Sec. \ref{p_distr}) of
fluctuations is studied by solving analytically the master equation for the microscopic
probabilities $p(n,t)$ and inferring the time-evolution of the global probability
$P(N,M,t)$. The phenomenon of {\it partial condensation} is also discussed.
Finally, in Sec. \ref{theconclusions} we briefly summarize our results and draw our conclusions.

\section{The statistical model} \label{themodel}

We consider $M$
independent random variables that take integer values $n_m=0,1,2\dots $ ($m=1,M$)
subject to a probability distribution $p(n_m)$, which in general depends on some parameters among
which, possibly, the time $t$.

The closely related problem where $n_m$ are continuous variables
behaves very similarly. In the following, in order to make the idea more concrete, we will speak
of $M$ {\it boxes} containing a total number
\be
N=\sum _{m=1}^M n_m
\ee
of {\it particles}, with an average value
$\langle N\rangle=\sum _{m=1}^M \langle n_m\rangle$,
where
$\langle n_m\rangle=\sum _nn\,p(n)$.

Particles can be exchanged with the external environment. Assuming
that the dynamics amounts to elementary moves where a single entity
can be added to or removed from a specific box,
the probability $p$ obeys the following master equation
\be
\frac{dp(n,t)}{dt}=-\left [w^+(n)+w^-(n)\right ]p(n,t)+
w^-(n+1,t)p(n+1,t)+w^+(n-1,t)p(n-1,t),
\label{master1}
\ee
where $w^+(n)$ is the transition rate to increase the number of particles,
i.e. $n\to n+1$, and $w^-(n)$ the one to decrease it, $n\to n-1$.
Here and in the following we denote by $p(n,t)$ the probability,
making the time-dependence explicit.
The master equation (\ref{master1}) is completely general for systems of discrete
variables  where
$N$ is not conserved as, e.g.,
spin models (Ising, Potts, Clock etc...) \cite{Corberi2010}.
We consider the following transition rates
\be
\left \{
\begin{array}{l}
w^+(n)=(n+2)^{-k} \\
w^-(n)=n^{-k}\left (1-\delta _{n,0}\right )
\end{array}
\right .
\label{trans_rate}
\ee
The Kronecker $\delta$-function guarantees that particles cannot be
extracted from an empty box.
The form (\ref{trans_rate}) is such that the evolution of a large cluster of $n$ particles located in a single box
is much slower than in the less populated ones, similarly to what happens in certain models of irreversible
aging processes \cite{Becker14}.
Notice that the transition rates (\ref{trans_rate}) obey detailed
balance:
\be
\frac{w^+(n-1)}{w^-(n)}=\frac{p_{st}(n)}{p_{st}(n-1)}\, , \,\,\,\,\,  \forall n>0 ,
\label{detbalanc}
\ee
where
\be
p_{st}(n)=\zeta^{-1}(k)(n+1)^{-k}
\label{p_stat}
\ee
and the normalization factor $\zeta (k)=\sum _{n=0}^\infty (n+1)^{-k}$ is the Riemann $\zeta$-function.
Let us stress that the more general choice for a transition rate obeying
the detailed balance condition (\ref{detbalanc}) is $w^+(n)=(n+2)^{-k}g(n+1)$,
$w^-(n)=n^{-k}\left (1-\delta _{n,0}\right ) g(n)$, where $g$ is an arbitrary function. Here we make the simplest
choice $g(n)\equiv 1$.

The form of $p_{st}$ above provides an average occupation
\be
\langle \rho \rangle =\langle n\rangle =\frac{\zeta(k-1)}{\zeta(k)}-1,
\label{averho}
\ee
a result that will be useful in the following. Notice also that $\langle \rho \rangle$
only exists for $k>2$ and, similarly, there is a finite variance only for $k>3$.

With the $w^\pm$ of Eq. (\ref{trans_rate})
the master equation (\ref{master1}) reads
\be
\left \{
\begin{array}{l}
\frac{dp(n,t)}{dt}=-\left [(n+2)^{-k}+n^{-k}\right ]p(n,t)+
(n+1)^{-k}\left [p(n+1,t)+p(n-1,t)\right ]\,,\,\,\,\,\, \forall n>0\\
\\
\frac{dp(0,t)}{dt}=-2^{-k}p(0,t)+p(1,t)
\end{array}
\right .
\label{master2}
\ee

The probability to have a total number $N$ of
particles at time $t$ is
\begin{eqnarray}
P(N,M,t)&=&\sum _{n_1,n_2,\dots,n_M}p(n_1,t)p(n_2,t)\cdots p(n_M,t)
\,\delta_{{\cal N},N} \nonumber \\
&=&\frac{1}{2\pi i}\oint dz \,e^{M[\ln Q(z,t)-\rho \ln z]},
\label{prob}
\end{eqnarray}
where ${\cal N}=\sum _{m=1}^Mn_m$, we have used the representation
$\delta _{{\cal N},N}=\frac{1}{2\pi i}\oint dz \, z^{-(N-{\cal N}+1)}$ and
\be
Q(z,t)=\sum_{n}p(n,t)z^{n}
\label{nq}
\ee
with $\rho =\frac{N+1}{M}\simeq \frac{N}{M}$ is the particle density.

The following relation
\be
P(N,M,t)=\sum _{n=0}^N \pi(n,N,M,t)
\label{recorP}
\ee
with
\be
\pi(n,N,M,t)=P(N-n,M-1,t)\,p(n,t),
\label{force}
\ee
is easily proved, see for instance Ref. \cite{bialas2}. $\pi(n,N,M,t)$ is the conditional probability that, 
at time $t$, there are $n$ particles in the $M$-th box,
given that a total number $N$ is found in all the boxes. Since the random variables are
identically distributed, the same probability applies to a generic box, not only to
the $M$-th.
The recursion (\ref{recorP}) allows one to determine the probability
distribution of $M$ variables from the one for $M-1$.
Specifically, once $p$ is known from the solution of the evolution
equations (\ref{master2}), Eqs. (\ref{recorP},\ref{force}) can be used
with the {\it boundary condition} $P(N,M=1,t)=p(n=N,t)$ to obtain
$P$, step by step, for larger and larger values of $M$.

It should be stressed that
Eq. (\ref{recorP}) makes the exact determination of $P$ feasible also for
reasonably large values of $N$ and $M$. Indeed, the computational
complexity using this formula is only polynomial, whereas
there is an exponential number of redundant operations involved
in the determination of $P$ by using the first line of Eq. (\ref{prob}).
This point is discussed in more detail in Appendix \ref{appA0}.

Besides Eq. (\ref{recorP}), which is always exact, for large $M$ one can alternatively determine
$P$ by evaluating the integral in
Eq. (\ref{prob}) by the method of steepest descent
\be
P(N,M,t)\simeq e^{-MR(\rho,t)},
\label{probsaddle}
\ee
where
\be
R(\rho,t)=-\ln Q\left [z^*(\rho,t)\right ]+\rho \ln z^*(\rho,t)
\label{probrate}
\ee
is the rate function and
$z^*$ is the value of $z$ for which the exponential argument
in Eq. (\ref{prob}) is maximum. This is provided by the following saddle-point equation
\be
z^*\frac{Q'(z^*,t)}{Q(z^*,t)}=\rho.
\label{saddle}
\ee
As we will see soon, however, a straightforward saddle-point evaluation of the
integral in Eq. (\ref{prob}) is not always doable.

In this paper the model introduced insofar is studied to understand
the basic mechanisms governing the occurrence of fluctuations, and the
mathematical structure behind. As already pointed out in the introduction, its
formulation is similar to other, physically
inspired and intensively studied models of Statistical Mechanics.

To begin with, collections of independent identically distributed random
variables obeying Eq. (\ref{p_stat}) have been introduced
as a simple description of quantum gravity \cite{bialas1,bialas2}.
This same model is also sometimes referred to as {\it urn} model,
or {\it balls and boxes} model. In this approach $N$
is an external control parameter. This means that -- at
variance with our analysis -- fluctuations of this quantity are forbidden
by construction.
What is usually studied in that context are, instead, the properties of the stationary state
as the control parameters $k$ and $N$
are varied.
The non-equilibrium dynamics
following an abrupt change of $k$ (playing the role of an inverse
temperature), has also been considered in \cite{godr}.
The evolution of the model in that case, however,
is ruled by a $N$-conserving stochastic equation
different from Eq. (\ref{master2}).

Another class of related problems are descriptions
of non-equilibrium driven systems with particles hopping on a lattice,
like the zero range process \cite{zrp}. In these systems the probability
at stationarity is factorized into single-site
distributions that, for particular choices of the hopping rates,
can take the form (\ref{p_stat}) \cite{zrp}. $N$ is a conserved
quantity also in these cases. Furthermore, the properties of independent random variables
distributed according to Eq. (\ref{p_stat}) have been discussed in relation to
a wealth of different physical situations, like in the notable case
of L\'evy walks \cite{levy}, and in other models \cite{others}.

\section{Stationary state} \label{stat}

Eq. (\ref{master2}) has the stationary solution (\ref{p_stat}).
The properties of the probability $P_{st}(N,M)$ in the stationary state have
been studied elsewhere \cite{corberijpa}.
Here we briefly mention some basic results
that will be needed in the following. We first derive them in a somewhat simplified
framework that provides physical hints to the mathematically more refined exposition that will 
be presented in the following Section.

\subsection{Simplified framework} \label{subseccond1}

A relatively simple description of the properties of the stationary state can be
obtained by considering the large-$k$ behavior.
In this limit $\langle \rho \rangle \to 0$. In fact, given the form of the
microscopic probabilities $p_{st}$ in Eq. (\ref{p_stat}),
as $k$ grows the chance of a non-vanishing
outcome $n$ becomes progressively smaller.
This result can be easily derived
from the exact expression (\ref{averho}).
Then, for large $k$, deviations with
$\langle \rho \rangle  \ll \rho \ll 1$ are possible. We will
focus our analysis in this range of densities.

Since the variables $n_m$ are identically distributed there is an obvious symmetry among
the boxes. Then, if such a symmetry is not spontaneously broken, the representative
configurations of the stationary state are expected to have the $N$ balls
fairly distributed among all the boxes. In this case, given that $\rho \ll 1$, most of them
will be empty and a comparatively smaller number, of order $N$, will contain
one ball. Given that $k$ is large, the chance for a single box to host more than one particle is very
small and will be neglected in the following.
The probability to have states with fairly distributed particles is
\be
P_{sym}(N,M)=p_{st}(1)^N \cdot p_{st}(0)^{M-N}
\cdot \Omega (N,M)= \zeta (k)^{-M} \cdot 2^{-kN} \cdot \Omega (N,M),
\label{psym}
\ee
where $\Omega (N,M)=\left ( \begin{array}{l} M\\N \end{array} \right )$
is the number of ways to choose the $N$ occupied sites out of the total $M$
and we have used Eq. (\ref{p_stat}).

We show now that, for large $M$, the probability $P_{sym}$ of this symmetric state
can be negligible compared to that of a {\it condensed} one,
where the symmetry among the boxes is broken and
a macroscopic number $n_c\propto M$ of particles is
accumulated in one of them.
The probability $P_{cond}$ of such a state is
\be
P_{cond}(N,M)= M\cdot  p_{st}(n_c)\cdot P_{sym}(N-n_c,M-1).
\label{pcond}
\ee
Here the first term, $M\cdot p_{st}(n_c)$, represents the probability
to place $n_c$ particles in the condensing box (the factor $M$ in front accounts for
the $M$ ways to choose it). The last term, $P_{sym}$, is the probability (given by Eq. (\ref{psym})) associated with the
remaining $N-n_c$, which are uniformly spread among the remaining $M-1$ boxes.

At large $M$, using the Stirling approximation for $\Omega$ and introducing
the density $\overline \rho$ of balls in non-condensing boxes through
$n_c=(\rho -\overline \rho)M$ one finds
\be
\frac{P_{sym}(N,M)}{P_{cond}(N,M)}\simeq
e^{-Mk\left \{(\rho-\overline \rho) \ln 2 -k^{-1}\left [s(\rho)-s(\overline \rho)\right ]\right \}},
\label{condwins}
\ee
where $s(x)=-x\ln x-(1-x)\ln (1-x)$ and sub-dominant terms have been dropped.
Eq. (\ref{condwins}) shows that the formation of the condensed state is
surely favored for $\rho >\overline \rho$,
because for large-$k$ the first term
in the argument of the exponential, namely the positive quantity
$(\rho-\overline \rho) \ln 2$, prevails over the second.
Similarly, the formation of the condensed phase is
unfavored for $\rho <\overline \rho$.

The discussion presented insofar is valid for large-$k$. However the basic results
apply also to the small-$k$ regime, provided that $k>2$. This will be shown
with the somewhat more refined calculation sketched in the
next subsection \ref{subseccond2}. We will also
identify $\overline \rho$ with the average value $\langle \rho \rangle$, given
in Eq. (\ref{averho}), and establish that condensation always occur, when $k>2$, for
$\rho >\langle \rho \rangle$.

\subsection{Some mathematical refinements}\label{subseccond2}

We now show how the results obtained with the simple approach of the preceding section
are confirmed by a more accurate treatment of the model equations where the condition
of large $k$ is released.
At stationarity and for large $M$ one has
\be
Q(z)= \zeta ^{-1}(k) \sum _{n=0}^\infty(n+1)^{-k}z^n
=\zeta ^{-1}(k)\,\frac{Li_k(z)}{z},
\ee
where $Li_k(z)$ is the polylogarithm (Jonqui\`ere's function),
and therefore
\be
zQ'(z)\simeq \zeta ^{-1}(k)\frac {Li_{k-1}(z)-Li_k(z)}{z}.
\ee
The saddle-point condition (\ref{saddle}) then reads
\be
\frac{Li_{k-1}(z^*)}{Li_k(z^*)}=\rho+1.
\label{sss}
\ee
For $k\le 2$ this equation always admits a solution and this corresponds to
the fact that condensation does not occur. In the following we will concentrate
on the sector with $k>2$. In this case Eq. (\ref{sss}) has solution
only in the region ${\cal I}$ with $\rho \le \langle \rho \rangle $
\cite{bialas1,bialas2,others,godr},
where $\langle \rho \rangle$ is given in Eq. (\ref{averho}).
In this range Eq. (\ref{probsaddle}) holds with
\be
R_{st}(\rho)=-\ln \left \{\zeta (k)^{-1}\frac{Li_k[z^*(\rho)]}{z^*(\rho)}\right \}+\rho \ln z^*(\rho)
\label{rst}
\ee
In the complementary sector $\backslash \hspace{-.2cm}{\cal I}$, that is
for $\rho >\langle \rho \rangle$, a straightforward saddle-point approach is not
available and the phenomenon of condensation occurs, namely a macroscopic
number $n_c\simeq N-\langle N\rangle$ of particles -- those that cannot
be accommodated in the {\it normal} state --  is accumulated in a single box,
as discussed in Sec. \ref{subseccond1}. In this case $P_{st}(N,M)$ is, for large $N$,
determined by the probability
$Mp_{st}(N-\langle N\rangle)$ that a single box contains such a huge amount of particles.
Comparing with Eq. (\ref{pcond}), this shows that
$\overline \rho = \langle  \rho \rangle$.
In summary, one finds the following behavior
\be
P_{st}(N,M)\sim\left \{ \begin{array}{ll}
e^{-MR_{st}(\rho)} & \mbox{for}\,\,\,\, \rho \le \langle \rho \rangle \\
 Mp_{st}(N-\langle N\rangle) & \mbox{for}\,\,\,\,\rho \gg \langle \rho \rangle,
\end{array}
\right .
\label{P_stat}
\ee
with $R_{st}$ given in Eq. (\ref{rst}).

\begin{figure}[h]
\vspace{2cm}
\centering
\rotatebox{0}{\resizebox{.85\textwidth}{!}{\includegraphics{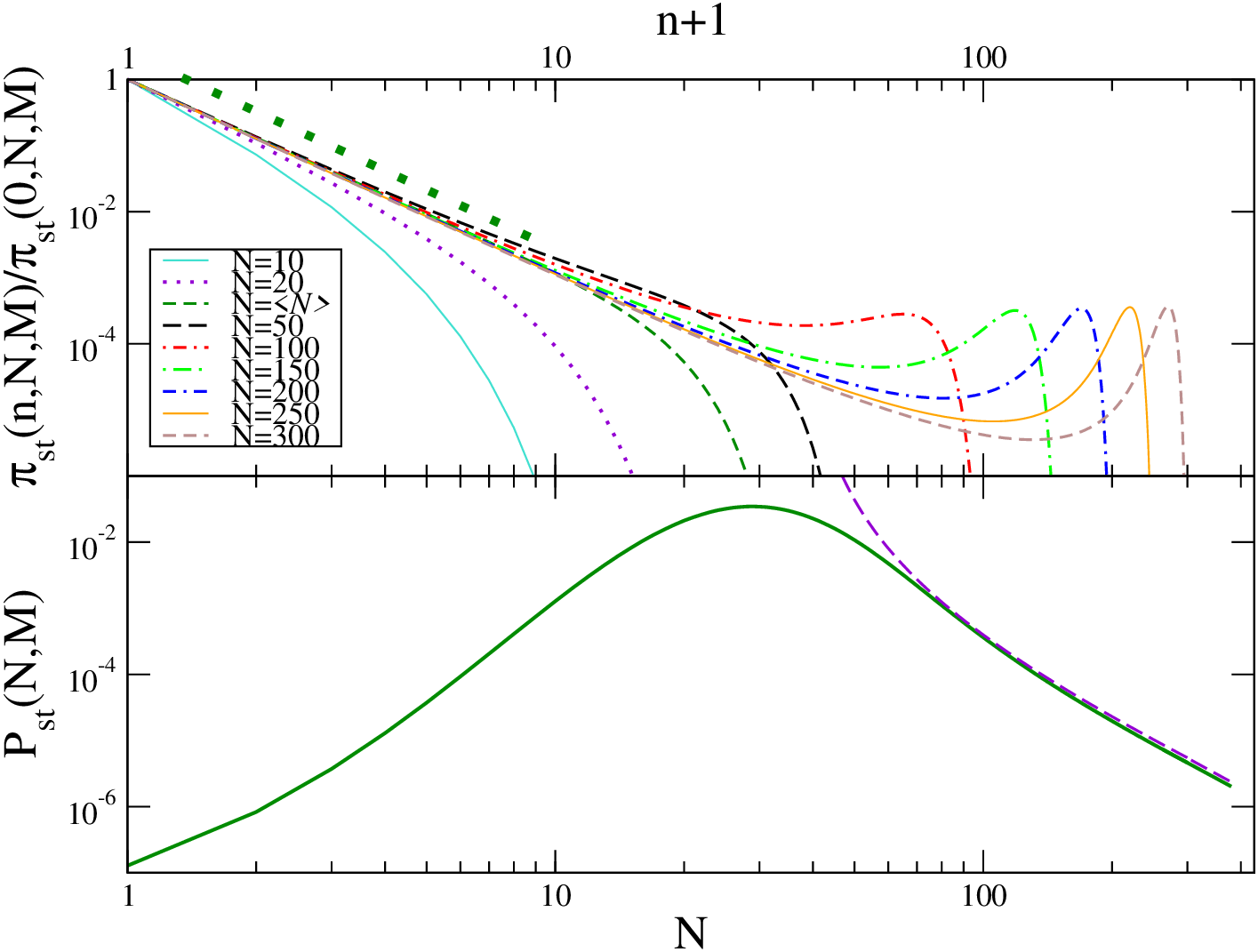}}}

\caption{In the upper panel the quantity $\frac{\pi_{st}(n,N,M)}{\pi_{st}(0,N,M)}$ of
Eq. (\ref{form_pi}) is plotted against $n+1$ in a double logarithmic scale,
for $M=100$ and different values of $N$ (see key). The dotted green-line is the
small-$n$ behavior $(n+1)^{-k}$. In the lower panel the
exact probability $P_{st}(N,M)$ with $k=3$ and $M=100$,
obtained by iteration of Eq. (\ref{recorP})
using the form (\ref{p_stat}) in Eq. (\ref{force}),
is plotted against $N$ with double logarithmic scales.
The dashed violet-curve
is the behavior $(N-\langle N\rangle)^{-3}$ of Eq. (\ref{P_stat}).}
\label{fig_static}
\end{figure}

The above expressions, covering the regions $ \rho \le \langle \rho \rangle$
and $\rho \gg \langle \rho \rangle$ only, can be derived analytically in the
large-$M$ limit.
The exact expression for $P_{st}$, which is valid for any $M$ and $\rho$
can instead only be obtained by iteration of the recursion
(\ref{recorP}), and is plotted in the lower half of Fig. \ref{fig_static}
(continuous green curve) considering the case with $k=3$ and $M=100$.
The curve has a maximum around $\langle N\rangle\simeq 36.84$.
We also compare the exact solution with the power-law behavior of Eq. (\ref{P_stat}) 
(dashed violet line), finding perfect agreement for
$N\gg \langle N\rangle$.

Eq. (\ref{P_stat}) shows that fluctuations with
$\rho \le \langle \rho \rangle$ behave
{\it normally}, in the sense that a large-deviation principle
with rate function $R_{st}$ is obeyed. On the other hand,
fluctuations with $\rho >\langle \rho \rangle$ are peculiar since
$P_{st}$ is not exponentially suppressed in $M$ and relatively
large fluctuations are possible. Indeed, rewriting the second line of Eq. (\ref{P_stat})
in terms of the density as
\be
P_{st}(N,M)\sim M^{1-k}\zeta^{-1}(k)(\rho-\langle \rho \rangle )^{-k} \hspace{1cm} \mbox{for}\,\,\,\,\rho \gg \langle \rho \rangle,
\label{tamedfluc}
\ee
one sees that, upon increasing $M$, a
fluctuation with
$\rho \gg \langle \rho \rangle$ is reduced only as $M^{1-k}$.
This gives a much better chance to detect large deviations with respect to a case where the LDP holds.

Condensation can be understood by
considering the conditional probability
$\pi_{st}(n,N,M)$ (Eq. (\ref{force}) at stationarity), which is shown
in the upper part of Fig. \ref{fig_static}. Here $\pi_{st}$ is plotted for $k=3$ and $M=100$,
and is normalized by
$\pi_{st}(n=0,N,M)$ in order to better
compare curves for different values of $N$. In this figure $\pi_{st}$ is obtained from Eq. (\ref{force}) by
evaluating $P_{st}$ by means of the recurrence (\ref{recorP}).

Let us discuss the properties of $\pi_{st}$.
Given its meaning, which is expressed below Eq. (\ref{force}),
it is clear that $\pi_{st} (n>N,N,M)=0$ in all cases, as it can be seen in 
the upper half of Fig. \ref{fig_static}.
Furthermore, by plugging Eq. (\ref{P_stat}) into Eq. (\ref{force})
one has
\be
\pi_{st} (n,N,M)\sim \left \{ \begin{array}{ll}
e^{-MR_{st}\left (\rho-\frac{n}{M}\right )}(n+1)^{-k} & \mbox{for} \,\,\,\,\, N-n \le \langle N \rangle \\
M^{1-k}\left (\rho -\langle \rho \rangle -\frac{n}{M}\right )^{-k}(n+1)^{-k} & \mbox{for} \,\,\,\,\,
N-n \gg \langle N\rangle,
\end{array}
\right .
\label{form_pi}
\ee
where we have replaced $M-1$ with $M$ and neglected
$\frac{1}{M}$ for large $M$.

This equation shows that, for fixed $N$ and $M$,
$\lim _{n\to 0}\pi_{st} (n,N,M)\propto (n+1)^{-k}$ for any value of $N$.
However, what makes the big difference between the normal and the condensed case
is the behavior of $\pi_{st}$ at large $n\lesssim N$.
Indeed, for $N\le \langle N \rangle $
a simple study of Eq. (\ref{form_pi}) (contained in Appendix \ref{appA1})
shows that $\pi _{st}(n,N,M)$ is a monotonically
decreasing function of $n$.
Hence, large values of $n$ are associated with a very small probability
$\pi_{st}$ and this implies that condensation -- namely a large fraction
of particles in a single box -- is probabilistically negligible.

This  can be checked in the upper panel of
Fig. \ref{fig_static}
The cases with $N\le \langle N\rangle$ discussed above
are represented by the curves with $N=10$ and
$N=20$, because using Eq. (\ref{averho}) with $k=3$ one finds
$\langle N\rangle\simeq 36.84$.
These curves decay monotonically, as expected.

Conversely, when condensation  occurs, there is an extensive
number $n_c$ of particles in a box, meaning that $\pi_{st}$ must be
non-negligible also for values of $n$ as large as $n=n_c$.
Indeed, for $N> \langle N \rangle $ (curves with $N\ge 50$),
$\pi_{st}(n,N,M)$ develops a pronounced maximum \cite{peak},
as it can be seen in Fig. \ref{fig_static} (upper panel) .
The location of the peak (see Appendix \ref{appA1}) is in
\be
n=n_c(N)= \eta \,(N-\langle N\rangle),
\label{nc1}
\ee
where $\eta \le 1$ is a function weakly dependent on $N$ and $M$
such that $\eta \to 1$ as $N$ or $M$ become large
(the dependence on $N$ can be checked by inspection
of Fig. \ref{fig_static}) .

The phenomenon of condensation of fluctuations is not restricted to the present model,
but it has been observed in a variety of different systems \cite{condfluc1,condfluc2,gamba},
not only related to Physics. Despite different in principle from the usual
condensation {\it on average} occurring in the prototypical example of a boson gas
\cite{bose} and in many other systems\cite{godr,condave}, these two kind of
condensation are related, as explained in \cite{condfluc1},
and a strong mathematical similarity exists.

\section{Dynamics} \label{dynam}

In the stationary state, for large $M$, a typical observation of the system
will give a value $N_{obs}$ of $N$ very close to $\langle N \rangle$.
However, if one waits enough, starting from this initial state
a fluctuation with $N\neq \langle N \rangle$ will develop
spontaneously. The aim of this paper is to describe the properties of the
dynamical process associated with the formation of such large deviations.
This is particularly interesting for a fluctuation with
$N>\langle N\rangle$, since in this case a macroscopic number $n_c$ of particles
must pile up in a single box whose occupancy was initially very small, and this
might be a slow and complex phenomenon.
Furthermore, once such a fluctuation sets on, it must regress and
this implies once again the dislocation of a large number $n_c$ of particles.
These two processes will be studied in the following Sections.
In order to do that, we start at $t=0$ with the typical form
of the single-variable probabilities in a system where the value
$N=N_{obs}$ is observed \cite{nota}. Recalling the meaning of $\pi$
(Eq. (\ref{force}) and discussion below), this reads
\be
p(n,t=0)=\frac{\pi_{st}(n,N_{obs},M)}{P_{st}(N_{obs},M)}.
\label{p_in1}
\ee
For instance, we will consider in the following section
the case where $N_{obs}=\langle N\rangle$ is the most
likely value of $N$ in order to study how a large deviation forms.
Then we will consider the evolution of the $p$-s by means of
Eq. (\ref{master2}) and, from the knowledge of $p(n,t)$ at all times
we will derive the form of $P(N,M,t)$ and of $\pi(n,N,M,t)$ using
Eqs. (\ref{recorP},\ref{force}) and/or Eqs. (\ref{probsaddle},\ref{probrate}).

\subsection{Creation of a fluctuation}

\subsubsection{Evolution of the single-variable probabilities $p(n,t)$} \label{p_crea}

\begin{figure}[h]
\vspace{2cm}
\centering
\rotatebox{0}{\resizebox{.85\textwidth}{!}{\includegraphics{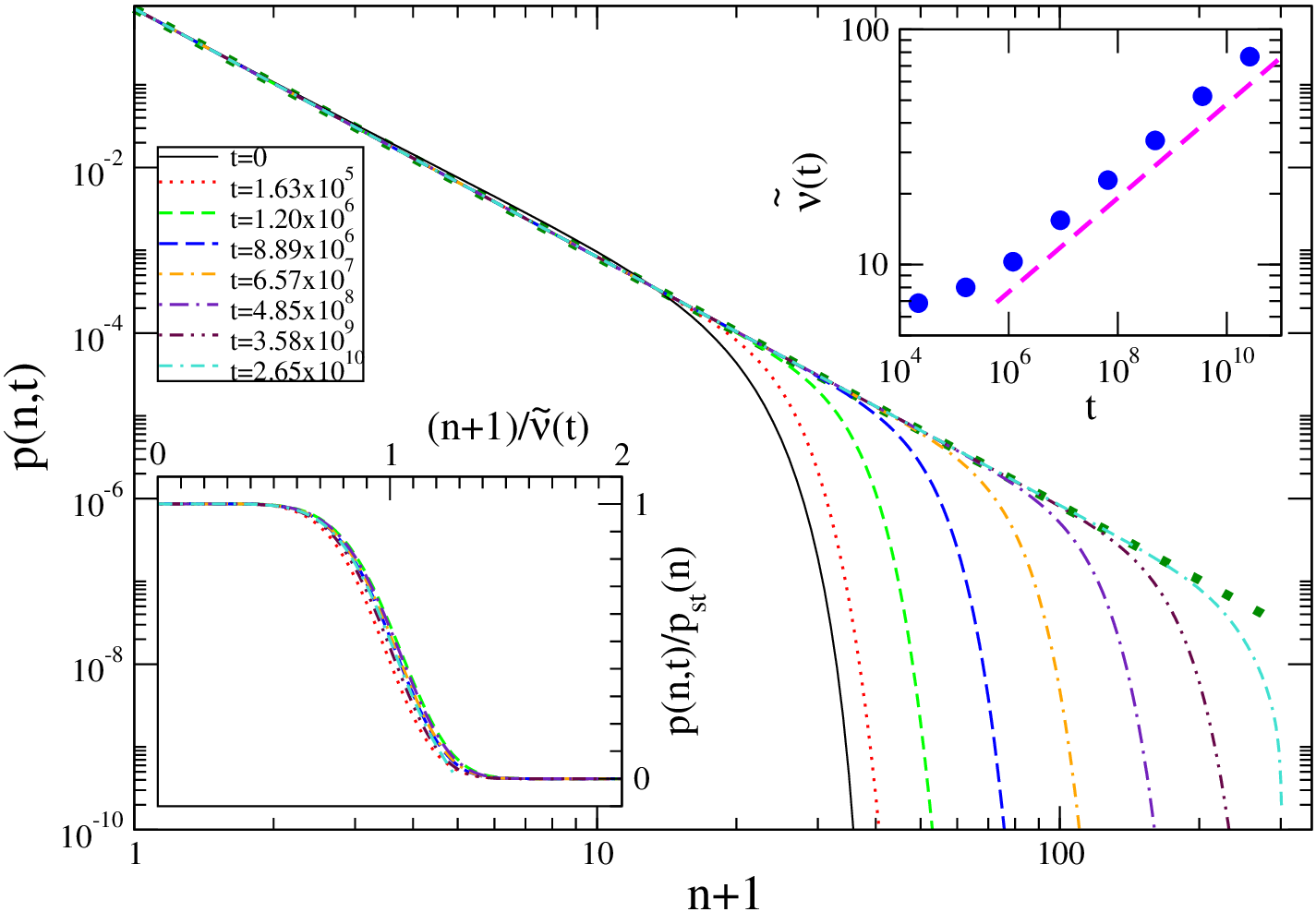}}}

\caption{The single-variable probability $p(n,t)$ with $k=3$
is plotted against $n+1$
with double logarithmic scales for different times (see key)
exponentially spaced.
The dotted green-line is the asymptotic form (\ref{p_stat}).
In the lower inset data collapse is tested by plotting
$\frac{p(n,t)}{p_{st}(n)}$ against $\frac{n+1}{\tilde \nu (t)}$.
The quantity $\tilde \nu (t)$ is plotted against
$t$ in the upper inset, in a double logarithmic plot.
The dashed magenta-line is the
expected long-time behavior (\ref{espz}).}
\label{fig_dyn_p_build}
\end{figure}

In this case we take $N_{obs}=\langle N\rangle$.
The dependence on $n$ of the initial condition (\ref{p_in1}) is
contained in the function $\pi_{st}(n,N_{obs}=\langle N\rangle,M)$.
As discussed regarding Eq. (\ref{form_pi}), this quantity
behaves as $p(n,0)\simeq (n+1)^{-k}$
for small $n$ and goes rapidly to zero for $n\simeq \langle N\rangle$.
This can be seen in the upper panel of Fig. \ref{fig_static}
(curve with $N=\langle N\rangle$) or
in Fig. \ref{fig_dyn_p_build} (leftmost black curve, for $t=0$, corresponding
to the initial condition (\ref{p_in1}) with $N_{obs}=\langle N\rangle$).
Hence we can write $p(n,0)\simeq p_{st}(n)f[(n+1)/\nu]$,
where $\nu \simeq \langle N \rangle$
and $f$ has the properties
\be
f(x)\simeq \left \{ \begin{array}{ll}
1 & \mbox{for}\,\,\,\, x\ll 1 \\
0 & \mbox{for} \,\,\,\, x\gg 1
\end{array}
\right .
\label{prop_f}
\ee
For long times one must approach the stationary condition (\ref{p_stat}).
Therefore, for large $t$, we search for a {\it scaling} solution of Eq. (\ref{master2}) of the form
\be
p(n,t)=\sigma(t)\,p_{st}(n)f\left [\frac{n+1}{\nu (t)}\right ]
\label{scal_p}
\ee
where $f(x)$ has the properties (\ref{prop_f}), $\nu $ is an
increasing function of $t$, and $\sigma $ is a weakly
time-dependent normalization such that
$\lim _{t\to \infty} \sigma (t)=1$.
Plugging this ansatz into the first line of Eq. (\ref{master2}) and performing
the calculations as detailed in Appendix \ref{appA} one can determine
the exact form of the scaling function $f(x)$ and of the
growth-law of $\nu$,
\be
\nu(t)=bt^{\frac{1}{k+2}},
\label{espz}
\ee
where $b$ is a constant.

The behavior of $p(n,t)$ as a function of $n$ at different
times, obtained by numerical integration of Eq. (\ref{master2}),
is shown in Fig. \ref{fig_dyn_p_build}. In the lower inset of this figure
we illustrate the data collapse obtained by plotting 
$\frac{p(n,t)}{p_{st}(n)}$ against $\frac{n+1}{\tilde \nu}$,
according to the scaling (\ref{scal_p}) (recalling that $\sigma (t)\simeq 1$).
The function $\tilde \nu (t)$ (eventually to be identified with $\nu (t)$) 
has been obtained looking for the best superposition of the curves at different times.
As shown in the upper inset, $\tilde \nu$ satisfies the behavior
Eq. (\ref{espz}) asymptotically. The collapse of the curves displayed in
the lower inset shows some correction at short times, progressively improving with increasing $t$.
Furthermore, data fall on a master curve that is almost
indistinguishable from the exact form of the scaling function $f(x)$ given in
Eq. (\ref{solf}) of Appendix \ref{appA}, when the undetermined parameter $a$ appearing in that
expression is appropriately tuned. This confirms the validity of our solution based on the scaling ansatz.

\begin{figure}[h]
\vspace{2cm}
\centering
\rotatebox{0}{\resizebox{.85\textwidth}{!}{\includegraphics{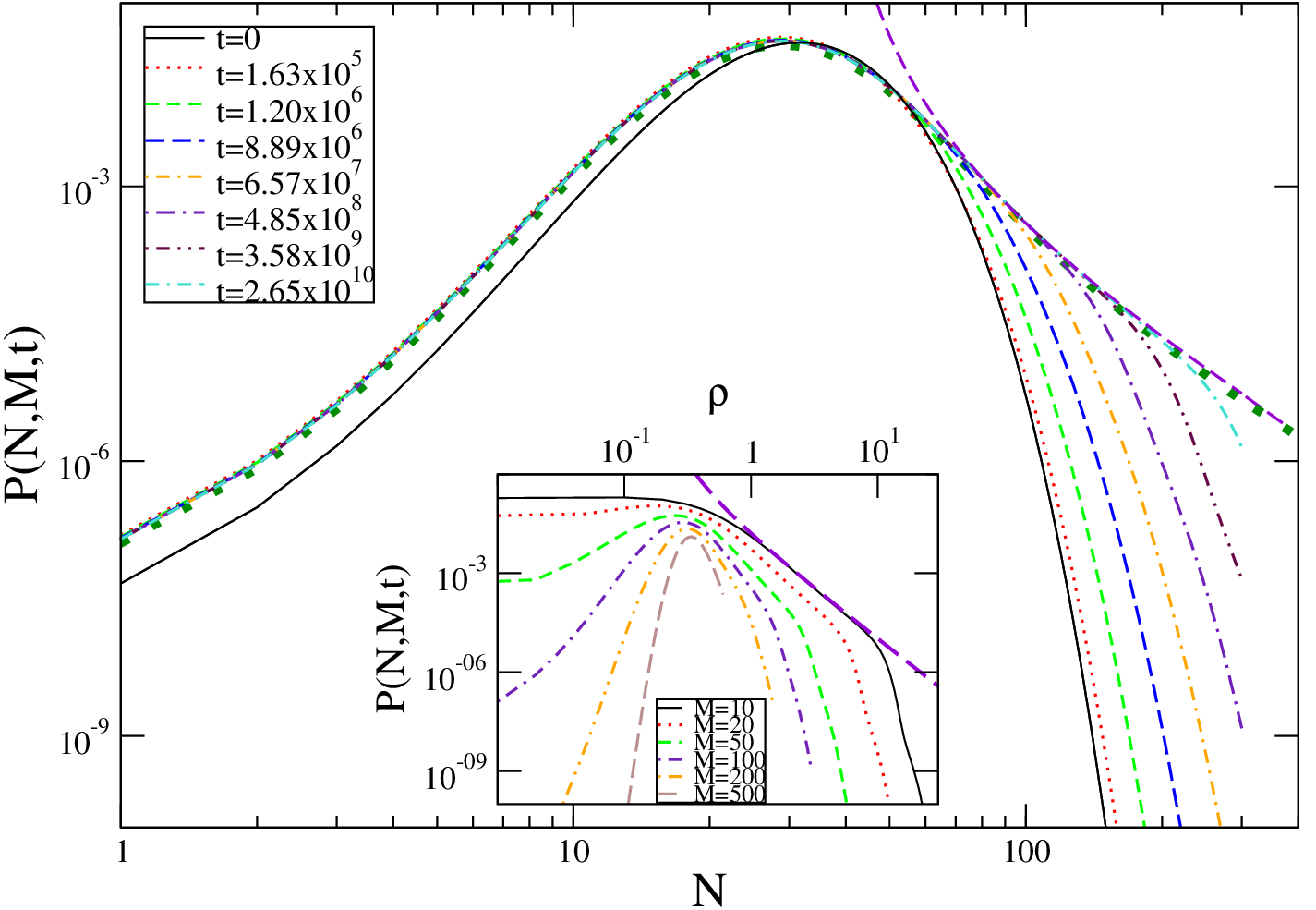}}}

\caption{The probability $P(N,M,t)$ with $k=3$ is plotted against $N$
with double logarithmic scales for different times (see key),
exponentially spaced.
The dotted green-line is the exact asymptotic form given by iteration of Eq. (\ref{recorP}),
using the form (\ref{p_stat}) in Eq. (\ref{force}) (it is the same curve as the green one
plotted in the lower panel of Fig. \ref{fig_static}).
The dashed violet-curve
is the behavior $(N-\langle N\rangle)^{-3}$ of Eq. (\ref{P_stat}). In the inset $P(N,M,t)$ is plotted
against $\rho$ for $t=4.85\cdot 10^8$ and different choices of $M$
($M=10,20,50,100,200,500$ starting from the curve on the top) on a double
logarithmic scale. The dashed violet-curve
is the behavior $(N-\langle N\rangle)^{-3}$.}
\label{fig_dyn_Pi_build}
\end{figure}

\subsubsection{Evolution of the collective probability $P(N,M,t)$} \label{sec_dyn_macro}

Once the form of the microscopic probabilities $p$ is found at all times, one can
obtain the exact time evolution of the probability $P$ of the collective quantity $N$
by inserting $p(n,t)$ in the recurrence equation (\ref{recorP}).
The outcome of this procedure is shown in Fig. \ref{fig_dyn_Pi_build},
where $P(N,M,t)$ is plotted against $N$ and different curves correspond to different
times (see caption). One observes that in the region $N<\langle N \rangle$, where
condensation does not occur, the asymptotic form $P_{st}$
(dotted green line) is reached already at an early stage. On the contrary, in the condensing part with
$N>\langle N \rangle$ the recovery towards $P_{st}$ is slow and proceeds gradually
from small to large values of $N$ as time goes on. In this way, at any time, no matter how long,
there exists a region of sufficiently large values of $N$ where stationarity is not
yet reached.

This behavior can be understood analytically. The analysis is presented in Appendix \ref{appB0}.
The main outcome of this study is that, at any time $t$, the dynamical probability $P(N,M,t)$
catches up with its stationary value $P_{st}(N,M)$ for densities
\be
\rho -\langle \rho \rangle\lesssim \frac{\nu}{M(k-1)}
\label{condrange}
\ee
larger then the average value. Recalling the discussion in Sec. \ref{stat},
this implies that the LDP is violated. Violation of the LDP in the dynamics 
occurs due to a mechanism analogous to the one operating at stationarity.
Indeed, also in the dynamical case, the steepest descent evaluation of
the integral in Eq. (\ref{prob}) cannot be done straightforwardly.
We also show that, for $\rho> \langle \rho \rangle$
outside the range (\ref{condrange}), the validity of the LDP is
restored, because the integral in Eq. (\ref{prob}) does admit a saddle point evaluation.
Clearly, this is trivially true in any case also for small densities
$\rho \le \langle \rho \rangle$.

Notice that the interval (\ref{condrange}) shrinks to zero as $M$ increases. Hence, for any finite
value of $\nu$, e.g. at any time, the validity of the steepest descent solution is recovered by
considering a number $M$ of boxes sufficiently large.
However, if $\nu =\infty$ (namely in the stationary state), the saddle point
evaluation fails for any $\rho > \langle \rho \rangle$ and condensation
occurs. Let us remark that the above
analysis implies that $\lim _{M\to \infty} \lim _{t\to \infty} P(N,M,t) \neq
\lim _{t\to \infty} \lim _{M\to \infty} P(N,M,t)$.

This interplay between $N$ (or equivalently $\rho$) and $M$ is shown in the inset of Fig. \ref{fig_dyn_Pi_build},
where $P$ is plotted against $\rho$ for $t=4.85\cdot 10^8$ (corresponding
to the third to last curve in the main picture), for different values of $M$.
Here it is clearly seen that, according to Eq. (\ref{condrange}), the region
of $\rho$ where $P$ coincides with $P_{st}$ shrinks as $M$ is increased until, at $M=500$ it is practically absent,
meaning that the LDP is recovered basically everywhere. 
The dashed violet line illustrates the large-$N$ behavior of Eq. (\ref{P_stat}).

It is clear that in the range of Eq. (\ref{condrange})
something akin to condensation
occurs, although its mathematical definition is less sharp than in
the stationary state, since we cannot let $M\to \infty$ because the interval (\ref{condrange})
would shrink to zero. This is supported by the observation that for ranges of $\rho $ increasing with time 
$P(N,M,t)$ becomes basically indistinguishable from $P_{st}(N,M)$ 
(see Fig. \ref{fig_dyn_Pi_build}). This implies
that in such ranges condensation occurs as at stationarity.
For larger values of $\rho$, however, $P$ departs from $P_{st}$, signaling that
condensation is absent. Since LDP is recovered, $P$ decays exponentially fast in $M$,
like in Eq. (\ref{ldp}), as opposed to the much softer algebraic decrease,
expressed by Eq. (\ref{tamedfluc}).
This explains why $P(N,M,t)$ drops off faster than
$P_{st}(N,M)$ (violet dashed line).

In order to understand the differences between the dynamical and the
stationary state, it is useful to consider
the conditional probability $\pi(n,N,M,t)$ of Eq. (\ref{force}).
We have evaluated this quantity at time $t=4.85\cdot 10^8$, which coincides with the
time at which we have plotted the third to last indigo curve in
Figs. \ref{fig_dyn_p_build}, \ref{fig_dyn_Pi_build}. 
From Fig. \ref{fig_dyn_p_build} one can infer that $\nu(t)\simeq 120$ at this particular time. 
Indeed one sees that $p(n,t)$ (indigo curve) is practically
identical to $p_{st}$ (dotted green line) up to this cutoff value of $n$, above which
$p$ decreases much more rapidly then $p_{st}$.

\begin{figure}[h]
\vspace{2cm}
\centering
\rotatebox{0}{\resizebox{.85\textwidth}{!}{\includegraphics{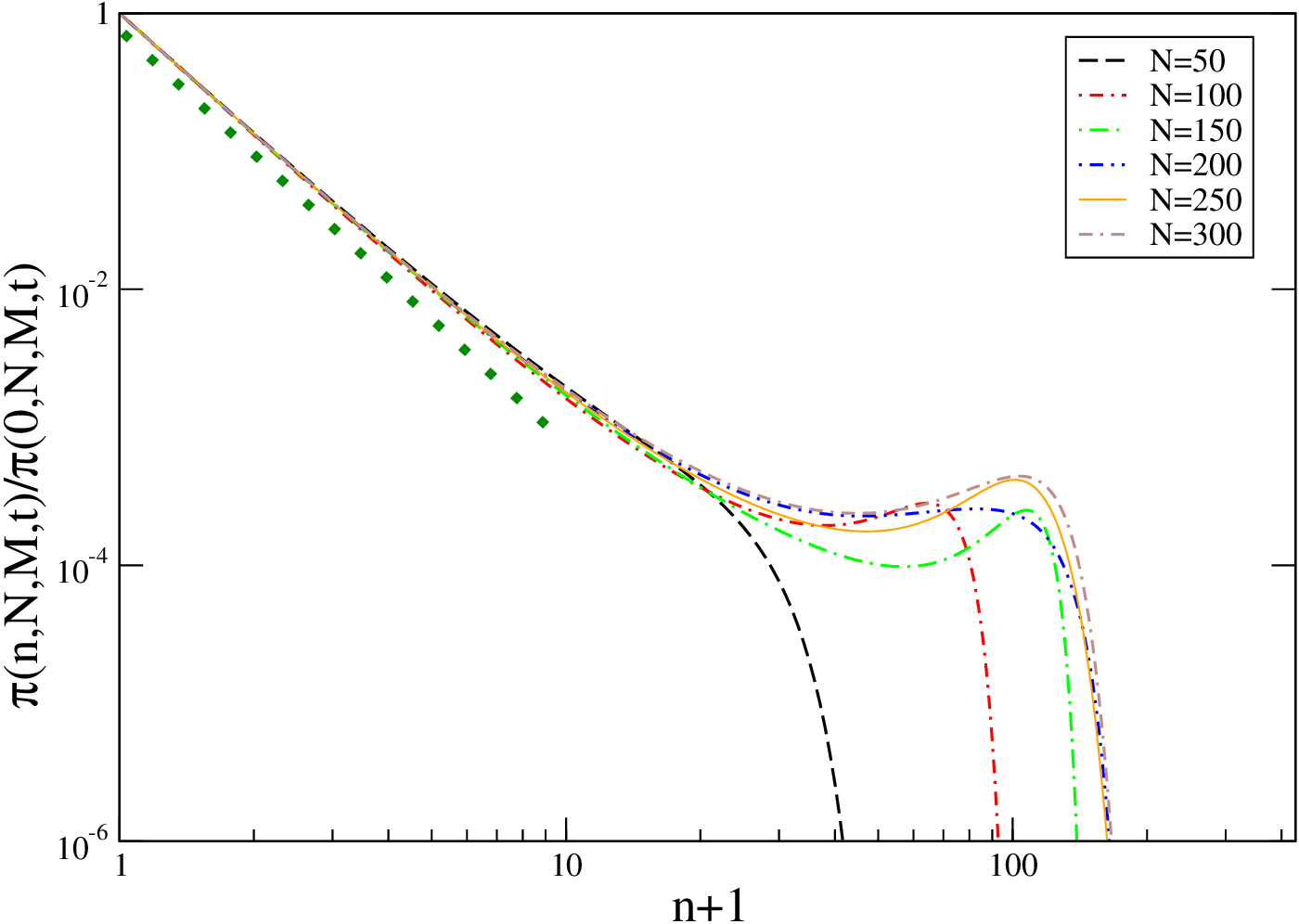}}}

\caption{The quantity $\frac{\pi(n,N,M,t)}{\pi(0,N,M,t)}$ for $t=4.85 \cdot 10^8$
of Eq. (\ref{form_pi}) is plotted against $n+1$ in a double logarithmic scale,
for $M=100$ and different values of $N$ (see key). The dotted green-line is the
small-$n$ behavior $(n+1)^{-k}$.}
\label{fig_pi}
\end{figure}

We have plotted $\pi(n,N,M,t)$ in Fig. \ref{fig_pi}
(normalized by $\pi (n=0,N,M,t)$). 
It is useful to contrast this probability with $\pi _{st}(n,N,M)$
plotted in the upper half of Fig. \ref{fig_static}.
Fig. \ref{fig_pi} shows that $\pi$ behaves similarly to $\pi _{st}$ for $N \lesssim \nu$: Upon increasing
$N$ a peak is developed around a value $n=n_c(N)$ (given in Eq. (\ref{nc1})) growing with $N$.
However, while for $\pi _{st}$ this continues to be true for any value of
$N$, no matter how large, the position of the relative maximum of $\pi $ saturates
around $n\simeq \nu$. This means that not all the $N-\langle N\rangle$
particles exceeding the average condense, but only a quantity of order $\nu$.
This {\it partial condensation} is obviously related to the fact that,
for $n>\nu$, the microscopic probability $p(n,t)$ rapidly vanishes and
the probability to condense more than $\nu$ balls is negligible.

In conclusion, at a given time $t$ the probability $P(N,M,t)$
has reached the stationary form $P_{st}(N,M)$ only up to
a value of $\rho$ given in Eq. (\ref{condrange}), while for larger values it is strongly suppressed.
Correspondingly, in the range (\ref{condrange}) a condensation phenomenon similar to the one observed at
stationarity is observed, with a number
$n_c(N)$ in Eq. (\ref{nc1}) of particles populating a
single box. For larger values of $N$, outside the interval (\ref{condrange}), only an incomplete condensation occurs
and a reduced number $n_c(\nu)<n_c(N)$ (with respect to what occurs
at stationarity) of particles is accumulated.
Notice that the approach of $P$ to $P_{st}$ is a slow, everlasting process,
since it is regulated by the power-law growth (\ref{espz}) of $\nu$.

\subsection{Regression of a fluctuation} \label{p_distr}

\subsubsection{Evolution of the single-variable probabilities $p(n,t)$}
 
\begin{figure}[h]
\vspace{2cm}
\centering
\rotatebox{0}{\resizebox{.85\textwidth}{!}{\includegraphics{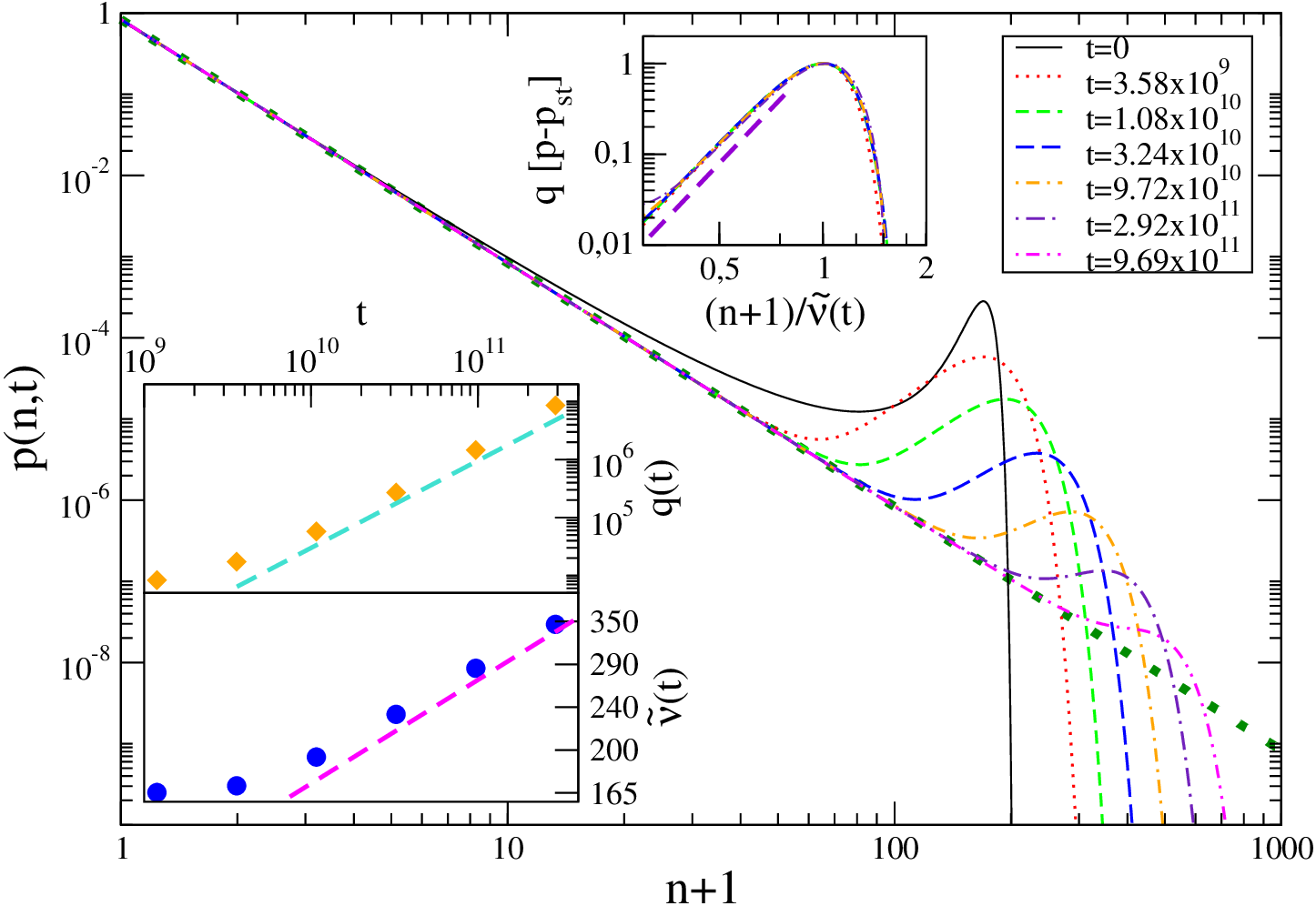}}}

\caption{The single-variable probability $p(n,t)$ with $k=3$ 
is plotted against $n+1$ 
with double logarithmic scales for 
different times (see key), 
exponentially spaced.
The dotted green-line is the asymptotic form (\ref{p_stat}).
In the upper inset data collapse is tested by plotting  
$q(t)[p(n,t)-p_{st}(n)]$ against $x=\frac{n+1}{\tilde \nu (t)}$.
The dashed indigo line is the behavior $x^4$.
In the lower inset the quantities $\tilde \nu (t)$ (below) and $q(t)$ 
(above) are plotted against $t$ in a double logarithmic plot. 
The dashed magenta-line (below) is the 
expected behavior (\ref{espz}) and the dashed turquoise-line (above)
is the behavior $\sim t^{3/2}$.}
\label{fig_dyn_p_destr}
\end{figure}

In order to study the process of the regression of a large fluctuation we take $N_{obs}\gg \langle N \rangle$.
According to Eq. (\ref{form_pi}) and the following discussion, 
the $n$-dependence of the initial condition (\ref{p_in1}) 
is given by the quantity $\pi_{st}(n,N_{obs},M)$. This behaves like
$p(n,0)\simeq (n+1)^{-k}$ for small $n$ and there is a peak at large $n$, 
centered around the value $n_c(N_{obs})$ of Eq. (\ref{nc1}), as shown in
the upper part of Fig. \ref{fig_static}. The initial
condition is represented in Fig. \ref{fig_dyn_p_destr} (curve for $t=0$, black). 
We can express these features with the following form
\be
p(n,0)\simeq p_{st}(n)f\left [\frac{(n+1)}{\nu}\right]+\nu^{-\alpha}g\left[\frac{(n+1)}{\nu}\right],
\ee 
where $\nu \simeq n_c(N_{obs})$, $f$ has the properties (\ref{prop_f}),
and $\nu^{-\alpha}$ and $g$ are an amplitude and a function
describing the behavior of the condensate (the peak).

Proceeding as in Sec. \ref{p_crea}, for large $t$ we search 
for a {\it scaling} solution of the form
\be
p(n,t)=\sigma(t)p_{st}(n)f\left [\frac{n+1}{\nu (t)}\right ]
+\nu ^{-\alpha}(t)g\left [\frac{n+1}{\nu (t)}\right ]
\label{scal_pd}
\ee
where, $\nu $, $\sigma $ and $f$ have the same meaning as in 
Sec. \ref{p_crea}, $\alpha $ is a dynamical exponent and
the scaling function $g$ has the following limiting behaviors
\be
\left \{ \begin{array}{l}
\lim _{x\to 0}g(x)=0 \\
\lim _{x\to \infty}g(x)=0.
\end{array} \right . 
\label{limf}
\end{equation}

Inserting the form (\ref{scal_pd}) into the first line of Eq. (\ref{master2})
and proceeding like in Appendix \ref{appA}
one has the time-dependence (\ref{espz}) of $\nu $ together with 
the form (\ref{solf}) of $f(x)$. The expression for 
$g(x)$ can also be determined. This is detailed in Appendix \ref{appB}.

The behavior of $p(n,t)$ as a function of $n$ at different 
times, obtained by numerical integration of Eq. (\ref{master2}), 
is shown in Fig. \ref{fig_dyn_p_destr}. 
According to Eq. (\ref{scal_pd}) a superposition of curves at
different times should be obtained by plotting 
$\nu ^\alpha \left \{p(n,t)-\sigma (t)p_{st}(n)f\left [\frac{n+1}{\nu}\right ]\right \}$ against $\frac{n+1}{\nu}$.
Given that $\sigma (t)\simeq 1$ for long times and
$f(x)\simeq 1$ for $x\lesssim 1$, data collapse can be checked 
by plotting $\nu ^\alpha \left \{p(n,t)-p_{st}(n)\right \}$ as well.

Since the value of $\alpha $ cannot be obtained from
the above calculation, 
we evaluate it from the numerical data as follows: 
The second term in Eq. (\ref{scal_pd})
describes the peak of $p$ observed in Fig. \ref{fig_dyn_p_destr}.
The amplitude $\nu ^{-\alpha}$ of such contribution at the peak
position $n_c(t)$ can then be estimated by measuring the difference 
$q^{-1}(t)=p(n_c,t)-\sigma (t)p_{st}(n_c)f\left (\frac{n_c+1}{\nu(t)}\right )$.
For large times it is $\sigma (t)\simeq 1$,
$n_c(t)\simeq \nu(t)$, and the first term in Eq. (\ref{scal_pd}) behaves as 
$p_{st}(n)$ up to $n\simeq \nu(t)\simeq n_c(t)$ (given the form of the function $f$,
see inset of Fig. \ref{fig_dyn_p_build}). 
Therefore the quantity $q^{-1}(t)$ can be approximately simplified to
$q^{-1}(t)\simeq p(n_c,t)-p_{st}(n_c)$.

In the upper inset of Fig. \ref{fig_dyn_p_destr} we plot
$q(t) \left \{p(n,t)-p_{st}(n)\right \}$
against $x=\frac{n+1}{\tilde \nu(t)}$,
where $\tilde \nu(t) $ is defined like in Sec. \ref{p_crea}.
As pointed out above,
data collapse of the curves at different times is expected in this 
plot in the region of the peak.
In this figure, in fact, an excellent superposition is found for long times. 
Notice that, 
using the small-$z$ behavior $L_n^\lambda(z)\simeq const.$ of
the Laguerre polynomials 
entering the form of $g$ (Eq. (\ref{scalfdis}), Appendix \ref{appB}), 
one has $g(x)\sim x^{k+1}$ for small-$x$, which is 
indeed very well observed in the upper inset of Fig. \ref{fig_dyn_p_destr}.
The behavior of $\nu (t)$ and $q(t)\sim \nu (t)^\alpha \sim t^{\frac{\alpha}{k+2}}$ 
as time changes is shown in the 
lower inset of the figure. This plot confirms the growth-law 
(\ref{espz}) of $\nu $ and indicates a value of $\alpha $ consistent
with $\alpha =15/2$ for the
case considered here with $k=3$.

\subsubsection{Evolution of the collective probability $P(N,M,t)$} \label{sec_dyn_macro2}

Eq. (\ref{scal_pd}) shows that the single-variable probability is roughly
the one at stationarity with a cut-off at $n+1\simeq \nu(t)$ and an extra
contribution (the second term on the r.h.s.) concentrated around
$n+1\simeq \nu(t)$. The latter, which represents the condensed fraction, is
clearly visible as a {\it bump} in Fig. \ref{fig_dyn_p_destr}.
Given this form of $p(n,t)$ it is easy to show that the 
collective probability $P$ exhibits a series of maxima located in $N=N_\ell(t)$
($\ell=0,1,2\dots$) given by
\be
N_\ell(t)=\langle N\rangle+\ell\nu(t).
\label{pos_max}
\ee
Indeed, there is an obvious maximum of the probability in 
the average value, $N_0=\langle N\rangle $. Furthermore, since the $p$'s
have a large support around $\nu$, the situation where, besides 
the $\langle N\rangle $ particles distributed among the boxes,
there are $\nu $ marbles stored in a single box (or, in general, in $\ell$ boxes) is also largely probable,
thus giving a series of maxima located like in Eq. (\ref{pos_max}).

\begin{figure}[h]
\vspace{2cm}
\centering
\rotatebox{0}{\resizebox{.85\textwidth}{!}{\includegraphics{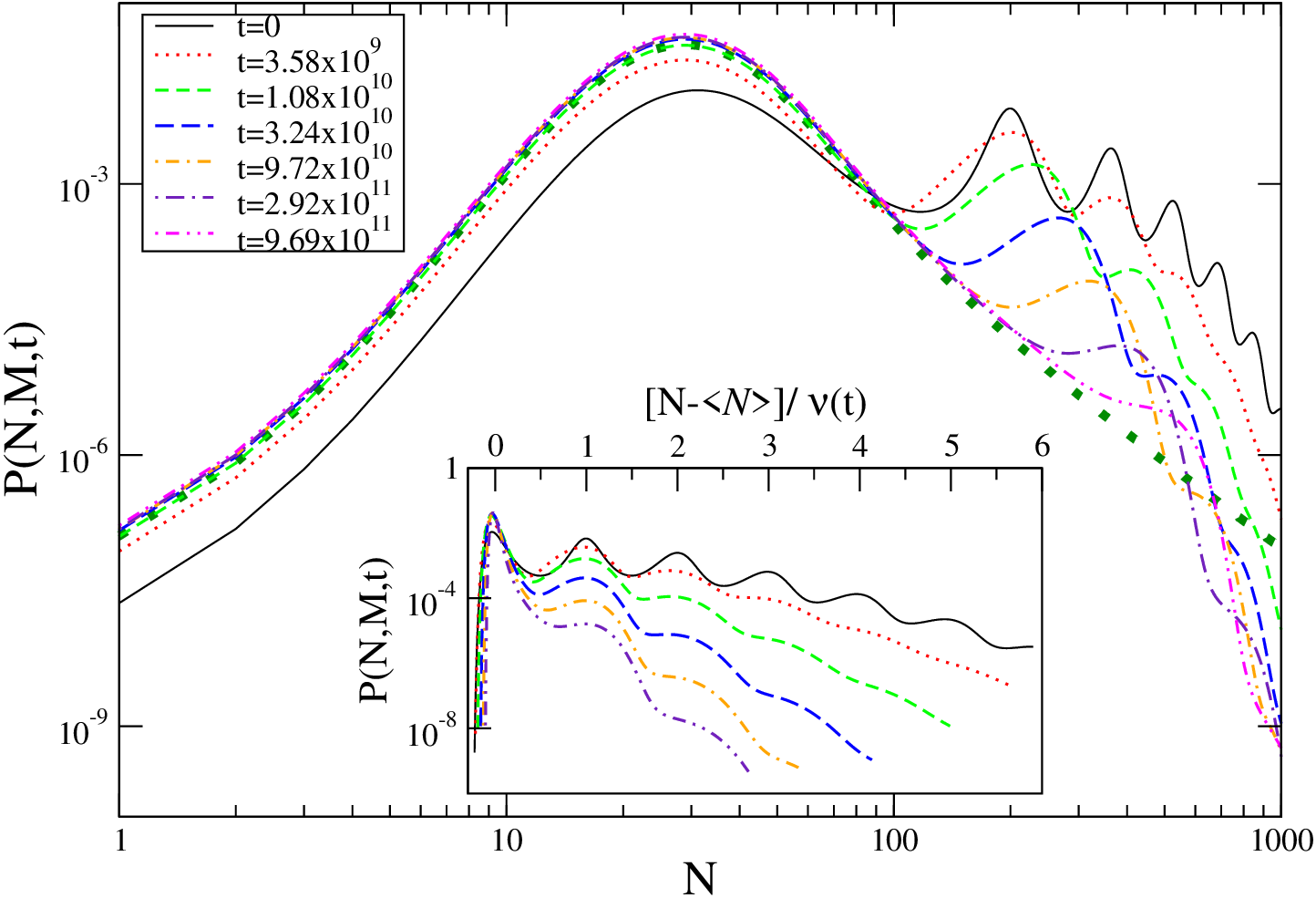}}}

\caption{The probability $P(N,M,t)$ with $k=3$ is plotted against $N$ 
(with double logarithmic scales) for different times (see key),  
exponentially spaced.
The dotted green-line is the asymptotic form (\ref{P_stat}).
In the inset the same quantity is plotted against 
$\frac{N-\langle N\rangle}{\nu (t)}$.}
\label{fig_dyn_Pi_destr}
\end{figure}

This structure of $P$ with many relative maxima is shown in 
Fig. \ref{fig_dyn_Pi_destr}. Also in this case $P$ has been computed by
inserting the time-dependent form of the $p$'s in Eq. (\ref{recorP}).
Clearly, as time goes on and $\nu $ increases
the relative {\it strength} of the condensed term decreases
(as $\nu ^{-\alpha}$) and the maxima are gradually smeared-out.
The location of the maxima (\ref{pos_max}) can be checked
by plotting $P$ against 
$\frac{N-\langle N\rangle}{\nu (t)}$, since on this axis the maxima are
placed on the integer values $\ell=0,1,2 \dots$. This is very neatly observed
in the inset of Fig. (\ref{fig_dyn_Pi_destr}).
From the discussion above it is clear that the maxima with $\ell>0$ are due
to the presence of the second term on the r.h.s. of Eq. (\ref{scal_pd}).
In the region $N<N_1(t)$, since the effect of this term is negligible, one 
recovers the asymptotic behavior $P(N,M,t)\simeq P_{st}(N,M)$ of 
Eq. (\ref{P_stat}), as it can be seen in Fig. \ref{fig_dyn_Pi_destr}.

By computing $\pi (n,N,M,t)$ one finds a behavior analogous to the 
one discussed in Sec. \ref{sec_dyn_macro}, signaling that also in this case
full condensation can only occur for sufficiently small values of $N$, whereas
only a partial one is possible for larger $N$. This 
is due to the same mechanism already discussed in Sec. \ref{sec_dyn_macro},
namely to the fact that the microscopic probabilities $p$ 
are negligibly small for $n>\nu$. 
Again, the approach of $P$ to the stationary form is an everlasting
slow evolution.

\section{Summary and conclusions} \label{theconclusions}

We have investigated the kinetics leading to the formation
or to the resorption of a large fluctuation of a collective
variable $N$ in a statistical system.
We have considered a simple model where $N$ is the sum
of a relatively large number $M$
of stochastic (micro) variables $n_m$ ($m=1,\dots ,M$)
identically and independently distributed. We speak of $n$ balls stored in
a box with probability $p(n)$, to make the idea more concrete.
The evolution equation of the $p$'s is
chosen as to have a stationary solution $p_{st}$ with
fat tails, namely $p_{st}(n) \propto (n+1)^{-k}$ ($k>2$).
It is known that this form induces a probability
$P_{st}(N,M)$ to observe a given value of $N$ at stationarity
that does not obey the LDP for
$N>\langle N\rangle$, due to a condensation
phenomenon. This feature implies that large fluctuations
in this range have a better chance to be observed in a system with finite but
large $M$ with respect to a case where the LDP holds. This
possibly makes some of our results amenable to
numerical and/or experimental verification.

We have considered the evolution of the model
starting from i) a typical situation where the most probable
value $\langle N\rangle$ is observed and
ii) a case where a measurement with unlikely outcome $N>\langle N\rangle$
just happened. In i) we follow the dynamical process whereby large
fluctuations, which are not present in the {\it average}
initial state, form.
Conversely, in ii) we consider a realization where
a rare initial state
is disrupted upon approaching stationarity.
Both these cases can be solved analytically. 
We have worked out the solutions in detail and reported them in the Appendices.

We have shown
that the convergence to stationarity is, in any case, a slow, everlasting
process akin to those observed in aging systems.
This happens because the evolution is slaved by the
power-law growth of a characteristic value $N=\nu (t)$
of $N$ separating a region $N<\nu(t)$, where the stationary
behavior has been attained, from one with $N>\nu (t)$,
where reminiscence of the initial condition is retained.
During such evolution one observes a condensation phenomenon
with a two-fold character: For $N\lesssim \nu(t)$ the phenomenon is
indistinguishable from the one observed at stationarity, with the
huge number $n_c(N)$ (\ref{nc1}) of particles stored in a single box.
For $N>\nu (t)$, instead, condensation is incomplete:
a relevant number of balls is still accumulated, but
this number only equals $n_c(\nu)$, which is smaller than
the value expected when full condensation occurs.
As time passes and $\nu (t)$ diverges, full condensation is
gradually recovered on increasing values of $N$.

Related to that, the interval $\backslash \hspace{-.2cm}{\cal I}$ where the LDP breaks down
has a non-trivial time-dependence.
We stress again that in this sector
large deviations may occur more easily.
To assess $\backslash \hspace{-.2cm}{\cal I}$ is therefore of practical
interest in nano-scale applications where fluctuations play an
important role.
In the stationary state, full condensation happens and this
gives $\backslash \hspace{-.2cm}{\cal I} \equiv \{\rho \vert \rho >\langle \rho \rangle \}$.
However, at finite times, since condensation is incomplete,
the LDP is spoiled only in the region given by
Eq. (\ref{condrange}). This means that, at any time, for a finite $M$
the LDP is restored for sufficiently large values of $\rho $,
at variance with what happens at stationarity.
The region (\ref{condrange}) expands as time elapses and $\nu (t)$ increases.
In this way the violation of the LDP progressively extends towards
the whole sector with $\rho >\langle \rho \rangle$, and the stationary
properties of $P$ are recovered.
Eq. (\ref{condrange})
shows that the size of $\backslash \hspace{-.2cm}{\cal I}$ can be tuned not only by changing $t$,
but also by acting on $M$.
Since $t$ and $M$ enter in the combination $\nu(t)/M$, this implies
that $t\to \infty$ and $M\to \infty$ are non-commuting limits.

The simple probabilistic setup discussed in this paper is suited to
describe at an elementary level the dynamics of fluctuations
in a variety of systems ranging
from Physics to Chemistry, Biology and Social Sciences.
It has the advantage of being amenable to analytic investigation.
Some of the general features displayed here
only rely on general aspects of probability, such as the violation of
the LDP.  Therefore, they are expected to be observed
with similar characteristics in a class of problems wider than the one
considered in this paper, e.g. with non-identically and/or non-independently distributed microscopic
variables. This makes the issue a
rather broad and general research topic worth of further investigations.
Finally, we remark that
collective variables defined differently
from the sum $N$, e.g. energy or heat fluxes
in solvable models of Statistical Mechanics, have been shown \cite{condfluc1} to display a
different condensation phenomenon, without violation of the LDP.
The question then arises of how the dynamics of fluctuations in these systems
compares with the case considered in the present paper. 
This investigation will be the subject of future work.

 \vspace{1cm}
 
 \noindent
 {\bf Acknowkledgements}
 
 \vspace{.5cm}
 
 \noindent
We thank F.Illuminati for a critical reading of the manuscript.

\appendix

\section{Computational complexity of Eqs. (\ref{recorP}) and (\ref{prob})} \label{appA0}

Suppose we want to determine $P$ by means of Eq. (\ref{recorP}).
Let us consider the procedure at
a certain step when $P$ is known for a number $m-1$ of boxes and
the task is to determine it for $m$ of them.
Eq. (\ref{recorP}) informs us that, in order to find $P(N,m,t)$, we must
preliminarily know $\pi(n,N,m,t)$ for any value of $n$.
It is obvious that in order to 
find $\pi(n,N,m,t)$ (for any $n$) by means of Eq. (\ref{force}) the 
value of $P(j,m-1,t)$ must be previously known for any 
$j=0,1,\dots,N$. This means that at any step  
$\pi(n,i,m-1,t)$ must be determined through
Eq. (\ref{force}) for any $n=0,1,\dots,N$ and $i=0,1,\dots,N$.
This requires $(N+1)^2$ computations. Once $\pi(n,i,m,t)$ is known
in this way,
we can get $P(j,m,t)$ through Eq. (\ref{recorP}) with further 
$(N+1)^2$ computations, since we have to sum up $N+1$ terms and this
operation must be repeated for any $j=0,1,\dots,N$.  
Then, for any step (namely going from $m-1$ to $m$) a number
of order $2(N+1)^2$ of elementary computations is needed. Since
the recurrence must be repeated up to $m=M$, a total number
$2(N+1)^2M$ of such calculations is needed. This must be compared
with the exponentially large number $(N+1)^M$ of operations involved 
in the determination of $P$ by using the first line of Eq. (\ref{prob}).

\section{Properties of $\pi _{st}$} \label{appA1}

For $N\le \langle N \rangle $ the first line of Eq. (\ref{form_pi}) necessarily
applies, which shows that not only $(n+1)^{-k}$ decreases upon increasing $n$ but also 
$e^{-MR_{st}\left (\rho-\frac{n}{M}\right )}=P_{st}(N-n,M)$. Referring to Fig. \ref{fig_static}
(lower part), this can be understood as follows:
In the first line of Eq. (\ref{form_pi}), $P_{st}(x,M)$ is evaluated for $x=N-n$.
Recalling that the condition $N-n \le \langle N\rangle $ applies,
this value of $x$ is located on the left of the maximum of $P_{st}$ located in $x=\langle N\rangle$ (or, at most,
on the maximum itself). Consequently,   
raising $n$ moves the argument $x$ of $P_{st}(x,M)$ further and further
away on the left of the maximum
(which amounts to descend
towards the left along  
the green-curve of the lower panel of Fig. \ref{fig_static}). This makes
$P_{st}(x,M)$ to decrease monotonically. The quantity  $e^{-MR_{st}\left (\rho-\frac{n}{M}\right )}$ in 
the first
line of Eq. (\ref{form_pi}) behaves similarly.
Then, large values of $n$ are associated with a very small probability
$\pi_{st}$ and this implies that condensation -- namely a large fraction
of particles in a single box -- is probabilistically negligible.

Conversely, for $N\gg \langle N\rangle$ and values of $n$ such that 
the lower row of Eq. (\ref{form_pi}) applies [i.e. for $N-n\gg \langle N\rangle$], 
while increasing $n$ the term $(n+1)^{-k}$ 
decreases,
the factor $\left (\rho -\langle \rho \rangle -\frac{n}{M}\right )^{-k}$ increases. The effect of this is the development of 
a pronounced maximum \cite{peak} in $\pi_{st}(n,N,M)$,
as it can be seen in Fig. \ref{fig_static} (upper panel) for $N>\langle N\rangle$. 
There is therefore a relatively high
probability of having a macroscopic -- namely of order $N$ -- number $n_c$ of particles 
condensed in a single box.

Using the second line of Eq. (\ref{form_pi}) the location
of the maximum is at
\be
n=n_c(N)= \frac{1}{2}\,(N-\langle N\rangle), 
\label{nc_a}
\ee
where $e=1/2$ is a constant. 
Notice that the simple calculation presented above to determine $n_c$ is not exact,
since the second line of Eq. (\ref{form_pi}) is only accurate for 
$N-n \gg \langle N\rangle$ while $n_c$ is located outside this range.
It can be shown however that the result (\ref{nc_a}) is basically correct, 
since the true behavior, expressed by Eq. (\ref{nc1}), only differs by the value of 
the prefactor. 

\section{Creation of a fluctuation:
Solution of the equation for the evolution of the the $p$'s} \label{appA}

Inserting Eq. (\ref{scal_p}) into the first line of Eq. (\ref{master2}) one arrives at
\begin{eqnarray}
\nu(t)^{k-1} \frac{d\nu(t)}{dt} x^{k+1} \left [\frac{\nu \frac{d\sigma}{dt}}{\sigma \frac{d\nu}{dt}}
\frac{f(x)}{x}-f'(x)\right ]&=& 
-\left [\left (1+\frac{1}{\nu x}\right )^{-k}+
\left (1-\frac{1}{\nu x}\right ) ^{-k}\right ]f(x)+ \nonumber \\
&+&\left [\left (1+\frac{1}{\nu x}\right )^{-k}f\left (x+\frac{1}{\nu}\right )+
\left (1-\frac{1}{\nu x}\right ) ^{-k} f\left (x-\frac{1}{\nu}\right )\right ]
\label{eqdiffdisc}
\end{eqnarray}
where $x=(n+1)/\nu $ like before, and $f'=df/dx$.
Now we make the {\it ansatz} that the quantity $\frac{\nu \frac{d\sigma}{dt}}{\sigma \frac{d\nu}{dt}}
$vanishes in the long-time limit. This will be checked for consistency at the end of the calculation.
In the same limit, when $\nu$ is large, we can 
expand the terms $(1\pm \frac{1}{\nu x})^{-k}\simeq 1\mp \frac{k}{\nu x}+\frac{k(k+1)}{2\nu ^2 x^2}$ and 
$f\left (x\pm \frac{1}{\nu}\right )\simeq f(x)\pm \frac{1}{\nu}f'(x)+
\frac {1}{2\nu ^2} f''(x)$ to 
second order in the small quantity $1/(\nu x)$ and retaining the leading terms one obtains
\be
-\nu(t)^{k+1} \frac{d\nu(t)}{dt} x^{k+1} f'(x)=
f''(x)-\frac{2k}{x}f'(x),
\label{eqdiff1}
\ee

Regarding Eq. (\ref{eqdiff1}) in the variables $x,t$, since the r.h.s. does not 
depend on $t$ one must have, on the l.h.s., $\nu ^{k+1} d\nu/dt =a$, where
$a>0$ is a constant. Hence
\be
\nu(t)=bt^{\frac{1}{k+2}},
\label{espzA}
\ee
where $b=[a(k+2)]^{\frac{1}{k+2}}$.
Eq. (\ref{eqdiff1}) then reads
\be
f''(x)+\left (ax^{k+1}-\frac{2k}{x}\right ) f'(x)=0,
\label{eqdiff2}
\ee
which, with the limiting behaviors (\ref{prop_f}), has the solution
\be
f'(x)=-cx^{2k}e^{-\frac{a}{k+2}x^{k+2}},
\ee
where $c>0$ is a constant. Integrating once again one arrives at
\be
f(x)=\frac{\Gamma \left (-\frac{3}{k+2},\frac{ax^{k+2}}{k+2}\right )}
{\Gamma  \left (-\frac{3}{k+2}\right )}
-\frac{u\left (1+\frac{k+2}{k^2+k-2}x^{k+2}\right )}{x^3}e^{-\frac{a}{k+2}x^{k+2}},
\label{solf}
\ee
where $\Gamma (\alpha)$ and $\Gamma (\alpha,y)$ are the $\Gamma$ and the 
incomplete $\Gamma$-function, and 
$u=\frac{\left [\frac{(k+2)^{k+5}}{a^3}\right ]^{\frac{1}{k+2}}}
{3\Gamma \left (-\frac{3}{k+2}\right )}$.
This form depends on the single parameter $a$, which is difficult to
determine since our solution is exact only asymptotically.
The quantity $\sigma (t)$ in Eq. (\ref{scal_p}) can be easily obtained
from the normalization of the probability as
\be
\sigma(t)-1\propto \nu(t) ^{1-k}.
\ee
Finally, from this equation and Eq. (\ref{espzA}) it is easy to verify the ansatz made 
after Eq. (\ref{eqdiffdisc}), namely that 
$\frac{\nu \frac{d\sigma}{dt}}{\sigma \frac{d\nu}{dt}} \to 0$ for $t\to \infty$. 

\section{Behavior of the collective probability $P$} \label{appB0}

Eq. (\ref{scal_p}) shows that the form of the single-variable probability $p$
is basically the one at stationarity with a cut-off at $n+1\simeq \nu (t)$.
We simplify the discussion about the evolution of $P$ by assuming that such cutoff is sharp. 
This amounts to 
approximate the actual behavior of $f$ 
given by Eq. (\ref{solf}) with the schematic form $f(x)=1-\theta (x-1)$. 
Reparametrizing time in terms of $\nu$ by means of Eq. (\ref{espz}), 
let us study the l.h.s. $S(z,\nu)=z\frac{Q'(z,\nu)}{Q(z,\nu)}$ of the saddle point equation (\ref{saddle}),
for arbitrary $z$. 
Since now
\be
Q(z,\nu)=\sum _{n=0}^\nu p_{st}(n)z^n=\zeta^{-1}(k)\sum _{n=0}^\nu(n+1)^{-k}z^n, 
\label{eqscuq}
\ee
one has
\be
S(z,\nu)=\frac{\sum _{n=0}^\nu n(n+1)^{-k}z^n}
{\sum _{n=0}^\nu(n+1)^{-k}z^n}.
\label{eqscut}
\ee 
Notice that, in the two equations above, the notation should be specified since
$n$ must run up to an integer number, say the closest to $\nu $.  
However this would not change the discussion below and we prefer to keep the simple notation of Eqs. (\ref{eqscuq},\ref{eqscut}).
As a function of $z$, $S$ rises steeply from $S(z=0,\nu)=0$ to an asymptotic value 
(for large $z$) that can be easily determined by retaining only the dominant term
with $n=\nu$ in the sums defining $S$ in Eq. (\ref{eqscut})
\be
\lim _{z\to \infty} S(z,\nu)=\nu.
\label{limitS}
\ee 
This asymptotic value is assumed for $z\gtrsim z_r$,
where $z_r$ can be evaluated as follows.
Let us consider the numerator on the r.h.s. of Eq. (\ref{eqscut}). 
For $z\gtrsim 1$, as a function of $n$,
the terms $n(n+1)^{-k}z^n$ decrease down to a value $n_r$ given by the largest
solution of the following equation 
\be
\ln z=\frac{k}{n_r+1}-\frac{1}{n_r}.
\label{zr}
\ee
For $n>n_r$, the argument of the sum in the numerator of Eq. (\ref{eqscut}) 
very rapidly diverge, because of the term $z^n$. 
A similar analysis for the denominator
shows that it behaves similarly, but with a slightly larger value
of $n_r$, that we denote by $n_{r,D}$, given by $\ln z=\frac{k}{n_{r,D}+1}$.  
Starting from $n_r=\infty$ when $z=1$, $n_r$ decreases upon raising $z$.
Notice that, if $z$ is too close to unity one has $n_r>\nu$, meaning that such 
value is not contained in the sums defining $S$ in Eq. (\ref{eqscut}).
Therefore a critical value $z_r$ exists
\be
\ln z_r=\frac{k}{\nu+1}-\frac{1}{\nu},
\label{steep}
\ee
such that, for $z>z_r$ this term starts to be contained in the sums 
in Eq. (\ref{eqscut}) and, beyond that value, these sums rapidly diverge.  
A  similar analysis carried out for the denominator
of Eq. (\ref{eqscut}) leads to a value, denoted by $z_{r,D}$, given  
by $\ln z_{r,D}=\frac{k}{\nu +1}$.
Since $z_r<z_{r,D}$, 
$S(z,\nu)$ is a 
very steep function for values of $z$ close to $z_r$, and then flattens when 
$z$ crosses also $z_{r,D}$ and the denominator diverges as well.
As Eq. (\ref{steep}) shows, $z_r\gtrsim 1$ for large $\nu$, 
which allows one to expand $\ln z_r$ around $z_r=1$ on the l.h.s. of Eq. (\ref{steep}) 
and to set $\nu +1\simeq \nu$ on the r.h.s., thus arriving at
\be
z_r=1+\frac{k-1}{\nu}.
\label{zN}
\ee 
This means that when the solution $z^*$ of Eq. (\ref{saddle}) 
moves the short distance from
$z^*=1$ to the nearby value $z^*=z_r$ given by Eq. (\ref{zN}), $\rho$ 
widely varies from $\rho =\langle \rho \rangle$ (according to Eq. (\ref{averho}))
to a value $\rho \sim \nu$ since, because of Eq. (\ref{limitS}),
the l.h.s. $S$ of the saddle point equation (\ref{saddle}) rapidly converges
to this value as soon as $z\gtrsim z_r$.

Let us now consider the argument $-M{\cal R}(z,\rho,\nu)$ of the exponential defining $P$ on the
r.h.s. of Eq. (\ref{prob}). Given that the largest contribution to the integral 
comes from $z\simeq z^*$ and, for
$\langle \rho \rangle \le \rho \lesssim \nu$, $z^*$ is close to $z^*=1$, 
in this range of $\rho$ one can write
\be
M{\cal R}(z,\rho,\nu )\simeq - M \left [\left . \frac{d\ln Q}{dz} \right \vert _{z=1}(z-1)-\rho (z-1)
\right ]
=M(\rho -\langle \rho \rangle)(z-1)
\label{exprate}
\ee
where we have expanded the argument of the exponential in Eq. (\ref{prob})
to first order in 
$z-1$ and we have used $\ln Q(1)=0$ (after Eq. (\ref{nq}) and normalization
of the probabilities $p$)
and $\left .\frac{d\ln Q}{dz}\right \vert _{z=1}=\langle \rho \rangle$
(from Eqs. (\ref{saddle},\ref{averho})).

Roughly speaking, for a given value of $M$,
a saddle-point evaluation of the integral defining $P$ in Eq. (\ref{prob})
is accurate if the positive quantity of Eq. (\ref{exprate}), as a function of $z$, has a pronounced
minimum at $z=z^*$. This in turn means that, whatever the value of $M{\cal R}$ 
at $z=z^*$ is, it must become much larger than its value for
other choices of $z$. 
However we know that, in the region $\rho \ge \langle \rho \rangle$
where condensation is possible, $z^*$ ranges at most up to $z^*=z_r$.
This implies that a saddle-point solution cannot be invoked if 
$M(\rho -\langle \rho \rangle)(z_r-1)$ is not a large number, namely if
$\rho -\langle \rho \rangle<[M(z-1)]^{-1}$.
Using Eq. (\ref{zN}), this means that the steepest descent evaluation
breaks down for all the densities
\be
\rho -\langle \rho \rangle\lesssim \frac{\nu}{M(k-1)}
\label{condrange_a}
\ee
larger but sufficiently near to the average value. Its validity is only
restored for larger values of $\rho$ (besides, clearly, for small densities
$\rho \le \langle \rho \rangle$).

\section{Regression of a fluctuation:
Solution of the equation for the evolution of the the $p$'s} \label{appB}

Inserting the form (\ref{scal_pd}) into the first line of Eq. (\ref{master2})
and proceeding like in Appendix \ref{appA}
one has the time-dependence (\ref{espz}) of $\nu $ and 
the form (\ref{solf}) of $f(x)$, whereas
for $g(x)$ one arrives at 
\be
g''(x)+ax^{k+1}g'(x)+\left [ a\alpha x^k-\frac{k(k+1)}{x^2}\right ]g(x)=0,
\label{eqdiff2d}
\ee
where $a=\nu^{k+1}d\nu/dt>0$ is the same constant introduced in Appendix \ref{appA}
(below Eq. \ref{eqdiff1}).
With the boundary conditions (\ref{limf}) the solution is
\be
g(x)=d\,x^{k+1} \, L_{-\frac{\alpha +k+1}{k+2}}^{\frac{2k+1}{k+2}}\left( -\frac{a}{k+2}x^{k+2}\right ),
\label{scalfdis}
\ee
where $d$ is a constant and $L_m^\lambda (z)$ are 
the generalized Laguerre polynomials.


\begin{thebibliography}{0}
\bibitem{nota}
Notice that our initial condition $p(n,t=0)$ is the typical form of the $p$'s when
the fluctuating variable $N$ takes the value $N=N_{obs}$,
but having the microscopic
probabilities set to $p(n,t=0)$ does not necessarily imply that the outcome of
a measurement of $N$ is $N=N_{obs}$, since obviously fixing
$p_{st}$ we do not constrain the value of $N$.
\end{thebibliography}
\end{document}